# Modification indices for diagnostic classification models

Christy Brown[a]* and Jonathan Templin[b]

[a]*Department of Education and Human Development, Clemson University, Clemson, SC, USA;*
[b]*Department of Psychological and Quantitative Foundations, University of Iowa, Iowa City, IA, USA*

*Correspondence concerning this article should be addressed to Christy Brown, Department of Education and Human Development, Clemson University, 210 Tillman Hall, Clemson, SC 29634; Office Phone: (864) 656-5121; Email: cjb2@clemson.edu

# Modification indices for diagnostic classification models

Diagnostic classification models (DCMs) are psychometric models for evaluating a student's mastery of the essential skills in a content domain based upon their responses to a set of test items. Currently, diagnostic model and/or Q-matrix misspecification is a known problem with limited avenues for remediation. To address this problem, this paper defines a one-sided score statistic that is a computationally efficient method for detecting under-specification at the item level of both the Q-matrix and the model parameters of the particular DCM chosen in an analysis. This method is analogous to the modification indices widely used in structural equation modeling. The results of a simulation study show the Type I error rate of modification indices for DCMs are acceptably close to the nominal significance level when the appropriate mixture $\chi^2$ reference distribution is used. The simulation results indicate that modification indices are very powerful in the detection of an under-specified Q-matrix and have ample power to detect the omission of model parameters in large samples or when the items are highly discriminating. An application of modification indices for DCMs to an analysis of response data from a large-scale administration of a diagnostic test demonstrates how they can be useful in diagnostic model refinement.

**Introduction**

Diagnostic classification models (DCMs; e.g., Rupp et al., 2010), also known as cognitive diagnosis models (e.g., Leighton & Gierl, 2007), are psychometric models allowing for the evaluation of an examinee's mastery of a set of predefined skills or attributes based upon their responses to a set of test items. The diagnostic approach to modeling item responses is in contrast to more traditional psychometric approaches that provide one overall measure of student ability in a general content domain. DCMs, however, provide estimates of student ability along multiple dimensions within a content domain. In education, this equips educators with an explanation for *why* a student is not performing well based upon the skills that have or have not yet been mastered, making it possible for educators to provide targeted remediation addressing individual strengths and weaknesses. Although DCMs have become an active area of research within the educational and psychological measurement community, many issues remain in assessing the fit of such models to empirical data.

The primary purpose in applying DCMs to a set of item response data is to classify examinees according to their degree of proficiency on multiple latent traits. However, if the statistical relation between attribute mastery and responses to the test items specified by the DCM used in the analysis is not correct, the resulting classifications will be questionable (e.g., Kunina-Habenicht et al., 2012; Rupp & Templin, 2008). Specification of a DCM includes two components: (1) identifying the latent attributes being measured by each item, typically conducted by subject-matter experts and then summarized as binary entries in what is known as a Q-matrix (Tatsuoka, 1983), and (2) defining the statistical model parameters that combine to predict item responses based upon the measured attributes an examinee has mastered. Therefore, any evaluation of diagnostic model fit should examine both the plausibility of the Q-matrix used in the analysis and the statistical parameters included in the model.

Research on diagnostic modeling has resulted in the development and refinement of many specific diagnostic model parameterizations such as the DINA, DINO, and C-RUM models, as well as the formulation of general diagnostic modeling families such as the log-linear cognitive diagnosis model (LCDM; Henson et al., 2009). For a thorough review of DCM parameterizations, see Rupp et al. (2010). Model fit can be assessed either in terms of the absolute fit, by quantifying how well responses predicted by the estimated model align with the observed responses, or in terms of the relative fit, by comparing the estimated model to an alternative one. However, measures of absolute fit do not provide an explanation for the source of the misfit. For this reason, we focus on measures of relative fit, which can identify specific sources of misfit and suggest potential pathways for improving model fit, in this study. Furthermore, diagnostic model fit can be examined at either the test or item level. Traditionally, relative fit at the test level has been carried out by assuming one model holds for all items and estimating the parameters of several specific DCMs, such as the DINA, DINO, and C-RUM, and then selecting the best fitting model according to measures of relative fit such as AIC or BIC (e.g., Chen et al., 2013). However, Henson et al. (2009) demonstrated how the most commonly used DCMs can each be represented using the LCDM formulation by placing statistical constraints on some of the model parameters, thus allowing the use of likelihood-based inference to test the statistical significance of the LCDM item parameters. That is, one could start by fitting a baseline model within the LCDM family, then add significant parameters and remove non-significant ones until the best fitting model is achieved. This may be the DINA for some items, the DINO for others, or a previously undefined DCM for other items. The study of Ma et al.

(2016) even found an improvement in examinee classification rate when different items were modeled by different DCM variants, as opposed to assuming a general model holds for all items. For these reasons, we will focus on inference-based measures of relative fit at the item level in this study.

Significance testing procedures for comparing nested models estimated under maximum likelihood include the likelihood ratio, Wald, and score tests. The likelihood ratio statistic requires estimation of parameters from both a full and reduced model and is thus the most computationally intensive of the three approaches. The Wald test requires estimation of the parameters in the full model only. For this reason, the Wald statistic is often used to test for model over-specification, that is, for removal of parameters currently in the model. The software package Mplus (Muthén & Muthén, 1998-2017) provides Wald statistics for LCDM item parameters (see Templin & Hoffman, 2013), and several studies have investigated the performance of Wald statistics in the DCM framework (e.g., de la Torre & Lee, 2013; Ma et al., 2016). The score (or Lagrange Multiplier) test requires only the estimation of the reduced model parameters. This makes the score statistic a computationally efficient test for model under-specification, that is, for testing whether the addition of certain model parameters would significantly improve model fit. The modification index widely used in structural equation modeling is in fact a one degree of freedom score statistic (Sörbom, 1989), and score tests have also previously been discussed in the context of item response theory modeling (Glas, 1999; Glas & Suárez-Falcón, 2003; Glas & Verhelst, 1995). Sorrel et al. (2017) investigated the use of the score test for the detection of over-specification in diagnostic models, which requires estimation of both the general and reduced DCM and is thus not a computationally efficient approach. Furthermore, the simulation study in Sorrel et al. (2017) found severely inflated Type I error rates when using the score test in this manner (e.g., observed rates as high as .30 for a nominal significance level of .05), necessitating the use of a simulated null distribution to assess statistical significance. As elaborated on in what follows, this could in part be due to the fact that the monotonicity constraints in diagnostic modeling mean some of the parameters of interest will be placed at their lower bounds in the null hypothesis of the score test. In this study, we seek to build on the work of Sorrel et al. (2017) by defining and evaluating the performance of score statistics appropriate for the detection of diagnostic model under-specification, that is, for fitting a reduced DCM and testing whether the addition of item parameters from a more general model significantly improve model fit, and by incorporating theory on score tests in a constrained parameter space to achieve Type I error control. We use the term *diagnostic model modification indices* for these score statistics to highlight the fact that they are used specifically for the detection of model under-specification and to connect to practitioners who are familiar with the use of the score test for this purpose in structural equation modeling.

As previously discussed, in addition to possible misspecification of the diagnostic model, the Q-matrix can also potentially be misspecified. The studies of Rupp and Templin (2008) and Choi et al. (2010) both examined conditions in which a Q-matrix was under-specified (that is, some attributes measured by an item were not recorded as such) and conditions in which a Q-matrix was over-specified (that is, attributes identified as measured by an item were not in fact related to the item). Both cases of Q-matrix misspecification led to decreased accuracy in parameter estimation and examinee classification, with the study of Choi et al. (2010) finding Q-matrix under-specification to be particularly detrimental. Kunina-Habenicht et al. (2012) also found Q-matrix misspecification to adversely impact examinee classification accuracy. DeCarlo

(2011) discussed how a potential misspecification of the Q-matrix for the fraction subtraction data (Tatsuoka, 1990) has led to some counter-intuitive examinee classifications.

Numerous methods for detecting Q-matrix misspecification have been proposed in previous studies (e.g., Chiu, 2013; de la Torre & Chiu, 2016; Kunina-Habenicht et al., 2012, Liu et al., 2012; Yu & Cheng, 2020). These methods showed promising results but none are perfect: many are based on computationally intensive search algorithms, some apply only to specific DCM variants, and none are inference-based. Moreover, none apply methods that are as well-known and are well-studied from other fields as the methods we develop in this study. Specifically, we define how the score statistic can also be used as a computationally efficient, inference-based, and empirically driven method of detecting Q-matrix under-specification. Therefore, we use the term *Q-matrix modification index* for these statistics as they parallel the modification indices widely used in structural equation modeling and have the advantage of connecting with practitioners familiar with this approach.

We note the investigation of over-specification of the Q-matrix in the likelihood-based context is already provided in maximum likelihood estimated DCMs by both Wald tests (when estimating a so-called alternative model) or by likelihood ratio tests (when estimating both a null and alternative model, when the null model is nested within the alternative). Asymptotically, the Wald, score, and likelihood ratio tests provide similar results. In practice, however, understanding model under-specification is difficult as model fit statistics are not often phrased in the context of potential model parameters. Moreover, in the case of some DCMs (i.e., DINA and DINO), the limited nature of their parameters makes assumptions that can easily be investigated by using modification indices.

The main purpose of this study is to define and investigate the statistical properties of score statistics for Q-matrix modification and for diagnostic model modification within the LCDM framework. The next section provides the technical details underlying the LCDM. This is followed by a definition of the score test and an outline of how it can be applied to the problem of improving model-data fit for the LCDM. Results from a series of simulation studies designed to assess the utility of score statistics for modification of the LCDM are presented as is an empirical application of modification indices for DCMs.

**Overview of the LCDM**

The purpose of a DCM is to classify examinees according to the attributes that they have or have not yet mastered. If a test measures $A$ dichotomous attributes, all possible combinations of mastery/non-mastery result in $2^A$ possible mastery profiles. Classification into these $2^A$ possible attribute mastery profiles is equivalent to assigning examinees to the appropriate *latent class* in a constrained latent class model (e.g., Lazarsfeld & Henry, 1968).

To formulate the general latent class model, let $Y_{ei}$ denote the response of examinee $e$ to item $i$ with $Y_{ei} = 1$ for a correct response and $Y_{ei} = 0$ for an incorrect response, $e = 1, 2, \ldots, E$ and $i = 1, 2, \ldots, I$. The latent class model assumes that the conditional distribution of $Y_{ei}$ given that examinee $e$ belongs to latent class $c$ is Bernoulli, with $\pi_{ic}$ representing the probability an examinee in latent class $c$ answers item $i$ correctly for $c = 1, \ldots, C$. Let $v_c$ denote the probability that a randomly selected examinee belongs to latent class $c$ with $\sum_{c=1}^{C} v_c = 1$. Then, the unconstrained latent class model defines the probability of observing a particular item response vector $\boldsymbol{y}_e = (y_{e1}, y_{e2}, \ldots, y_{eI})$ for examinee $e$ as:

$$P(Y_e = y_e) = \sum_{c=1}^{C} v_c \prod_{i=1}^{I} \pi_{ic}^{y_{ei}} (1 - \pi_{ic})^{1-y_{ei}}. \qquad (1)$$

The class membership probabilities $v_c$ are referred to as structural parameters, and the summation portion of the model is referred to as the *structural component*. The product across items stems from the local independence assumption and is referred to as the *measurement component* of the model. The latent class model parameters can be estimated using the method of maximum likelihood (see Bartholomew & Knott, 1999, Chapter 6). The different DCM variants can be formulated by placing appropriate constraints on the item response probability parameters $\pi_{ic}$ in the measurement component of the model.

The LCDM item response function is determined in part by the attributes being measured by each item. Attributes enter the model as categorical latent variables, with the attribute mastery profile for an examinee in latent class $c$ given by $\boldsymbol{\alpha}_c = (\alpha_{c1}, \alpha_{c2}, \ldots, \alpha_{cA})$ where $\alpha_{ca} = 1$ indicates mastery of attribute $a$ and $\alpha_{ca} = 0$ indicates non-mastery of the attribute. The item by attribute Q-matrix specifies which attributes are measured by each item, with $q_{ia} = 1$ indicating that item $i$ measures attribute $a$ and $q_{ia} = 0$ indicating that it does not.

As an example of the LCDM parameterization, consider an item $i$ that measures attributes 1 and 2 so that $q_{i1} = 1$ and $q_{i2} = 1$. Conditional on the attribute mastery profile $\boldsymbol{\alpha}_c$ for the latent class $c$ to which examinee $e$ belongs, the LCDM item response function for this item is:

$$\pi_{ic} = P(Y_{ei} = 1|\boldsymbol{\alpha}_c) = \frac{\exp(\lambda_{i,0} + \lambda_{i,1,(1)}\alpha_{c1} + \lambda_{i,1,(2)}\alpha_{c2} + \lambda_{i,2,(1,2)}\alpha_{c1}\alpha_{c2})}{1 + \exp(\lambda_{i,0} + \lambda_{i,1,(1)}\alpha_{c1} + \lambda_{i,1,(2)}\alpha_{c2} + \lambda_{i,2,(1,2)}\alpha_{c1}\alpha_{c2})}. \qquad (2)$$

Thus, the LCDM models the conditional item response probability via a logit link function with the linear predictor resembling a factorial ANOVA model where the measured attributes represent fully crossed and reference coded design factors. The first subscript on the $\lambda$ parameters refers to the item, the second to the level of the effect (i.e., 0 for intercept, 1 for main effect, 2 for two-way interaction, etc.), and the parenthetical subscripts identify the attributes with which the parameter is associated.

Comparing the linear predictor in Equation 2 for examinees having mastered exactly one of the measured attributes (that is, either $\alpha_{c1} = 1$ and $\alpha_{c2} = 0$, or $\alpha_{c1} = 0$ and $\alpha_{c2} = 1$) to that for examinees not having mastered any measured attributes (that is, $\alpha_{c1} = 0$ and $\alpha_{c2} = 0$), we see that the following restrictions are needed to ensure that examinees having mastered one attribute have a higher probability of responding correctly than examinees not having mastered either attribute:

$$\lambda_{i,0} + \lambda_{i,1,(1)} > \lambda_{i,0} \implies \lambda_{i,1,(1)} > 0$$
$$\text{and } \lambda_{i,0} + \lambda_{i,1,(2)} > \lambda_{i,0} \implies \lambda_{i,1,(2)} > 0 \ . \tag{3}$$

In general, all LCDM main effects must be positive in order to ensure that masters of a given attribute have a higher probability of a correct response than do non-masters. For Equation 2, we see that the following order constraints for the two-way interaction are also necessary:

$$\lambda_{i,0} + \lambda_{i,1,(1)} + \lambda_{i,1,(2)} + \lambda_{i,2,(1,2)} > \lambda_{i,0} + \lambda_{i,1,(1)} \implies \lambda_{i,2,(1,2)} > -\lambda_{i,1,(2)}$$
$$\text{and } \lambda_{i,0} + \lambda_{i,1,(1)} + \lambda_{i,1,(2)} + \lambda_{i,2,(1,2)} > \lambda_{i,0} + \lambda_{i,1,(2)} \implies \lambda_{i,2,(1,2)} > -\lambda_{i,1,(1)} \ . \tag{4}$$

The LCDM can be expressed in a general form as:

$$\pi_{ic} = P(Y_{ei} = 1 | \boldsymbol{\alpha}_c) = \frac{\exp\left(\lambda_{i,0} + \boldsymbol{\lambda}_i^T \boldsymbol{h}(\boldsymbol{\alpha}_c, \boldsymbol{q}_i)\right)}{1 + \exp\left(\lambda_{i,0} + \boldsymbol{\lambda}_i^T \boldsymbol{h}(\boldsymbol{\alpha}_c, \boldsymbol{q}_i)\right)} \tag{5}$$

where $\boldsymbol{\lambda}_i$ is a column vector containing the $2^A - 1$ main effect and interaction terms for item $i$ and $\boldsymbol{q}_i$ is the $i^{th}$ row of the Q-matrix indicating the attributes measured by item $i$. The column vector $\boldsymbol{h}(\boldsymbol{\alpha}_c, \boldsymbol{q}_i)$ contains linear combinations of $\boldsymbol{\alpha}_c$ and $\boldsymbol{q}_i$ such that:

$$\boldsymbol{\lambda}_i^T \boldsymbol{h}(\boldsymbol{\alpha}_c, \boldsymbol{q}_i) = \sum_{a=1}^{A} \lambda_{i,1,(a)} \alpha_{ca} q_{ia} + \sum_{a=1}^{A-1} \sum_{a'=a+1}^{A} \lambda_{i,2,(a,a')} \alpha_{ca} \alpha_{ca'} q_{ia} q_{ia'} + \cdots . \tag{6}$$

The first $A$ elements of the $\boldsymbol{\lambda}_i$ vector are the main effects for item $i$, the next $\binom{A}{2}$ are the two-way interactions, the next $\binom{A}{3}$ are the three-way interactions, and so on up until a final $A$-way interaction term for items measuring all $A$ attributes. Order constraints must also be imposed on these higher-order interaction terms to guarantee that the item response probability increases as additional attributes are mastered.

Constraints may also be placed upon the $\nu_c$ parameters in the structural component of Equation 1 through what is referred to as a structural model. By imposing constraints on the $\nu_c$ parameters, structural models reduce the number of parameters that need to be estimated. Several methods for modeling the structural parameters have been proposed in the DCM literature, including a log-linear model (Henson & Templin, 2005) and a structured tetrachoric model (de la Torre & Douglas, 2004; Templin, 2004).

The choice of models for both the $\nu_c$ and $\pi_{ic}$ parameters in Equation 1 completely specifies a diagnostic model. Estimation of these model parameters and calculation of $P(c|\boldsymbol{y}_e)$ leads to the classification of examinees into attribute mastery profiles, with classification made to the latent class for which an examinee has the highest membership probability. However, as discussed in the introduction, the accuracy of examinee classifications can be impacted if the DCM or its Q-matrix are misspecified. Thus, methods for detecting model misspecification are an important part of the model fitting process.

**Likelihood theory and score tests**

The modification indices for DCMs proposed in this paper are based upon the score test, a general hypothesis testing procedure useful in the detection of model under-specification. Thus, an overview of the score test for a general parametric model will be provided before describing how the score test can serve as an empirically driven method for modifying the parameters of a diagnostic model and its associated Q-matrix.

The score test considers the adequacy of a reduced (potentially under-specified) statistical model. That is, the fully-specified model contains $p$ parameters, but a model with only $p - q$ parameters is estimated; the $q$ remaining parameters have been fixed to zero in estimation and we would like to see if freely estimating these parameters would significantly improve the fit of the model to the data. As the score test only requires estimation of the $p - q$ parameters in the reduced model, it is often preferred over equivalent hypothesis testing procedures such as the likelihood ratio test, which requires estimation of both the full and reduced models, and the Wald test, which requires estimation of all $p$ parameters in the full model.

To define the score statistic, let $\boldsymbol{\beta}$ be a $p \times 1$ vector of model parameters and partition $\boldsymbol{\beta}$ as $\boldsymbol{\beta}^T = (\boldsymbol{\beta}_1^T, \boldsymbol{\beta}_2^T)$ where $\boldsymbol{\beta}_1$ is a $(p - q) \times 1$ vector of the nuisance parameters and $\boldsymbol{\beta}_2$ a $q \times 1$ vector of the parameters of interest in hypothesis testing. The adequacy of the reduced model is then tested by the null hypothesis $H_0: \boldsymbol{\beta}_2 = \boldsymbol{0}$. Let $\ell(\boldsymbol{\beta})$ denote the log-likelihood function of the model containing all $p$ parameters. Denote the score vector by $\boldsymbol{S}(\boldsymbol{\beta})$ and partition it as:

$$\boldsymbol{S}(\boldsymbol{\beta}) = \frac{\partial}{\partial \boldsymbol{\beta}} \ell(\boldsymbol{\beta}) = \begin{pmatrix} \frac{\partial}{\partial \boldsymbol{\beta}_1} \ell(\boldsymbol{\beta}) \\ \frac{\partial}{\partial \boldsymbol{\beta}_2} \ell(\boldsymbol{\beta}) \end{pmatrix} = \begin{pmatrix} \boldsymbol{S}_1(\boldsymbol{\beta}) \\ \boldsymbol{S}_2(\boldsymbol{\beta}) \end{pmatrix}. \tag{7}$$

Solving $\boldsymbol{S}_1(\boldsymbol{\beta}_1^T, \boldsymbol{0}^T)^T = \boldsymbol{0}$ gives the maximum likelihood estimates of the reduced model containing only $p - q$ parameters, $\widehat{\boldsymbol{\beta}}^T = (\widehat{\boldsymbol{\beta}}_1^T, \boldsymbol{0}^T)$. Let $\boldsymbol{I}_1(\boldsymbol{\beta})$ be the information matrix for a single observation and partition it according to the partitioning of $\boldsymbol{\beta}$:

$$\boldsymbol{I}_1(\boldsymbol{\beta}) = \begin{pmatrix} \boldsymbol{I}_{11} & \boldsymbol{I}_{12} \\ \boldsymbol{I}_{12}^T & \boldsymbol{I}_{22} \end{pmatrix} \tag{8}$$

where $\boldsymbol{I}_{11}$ is $(p - q) \times (p - q)$, $\boldsymbol{I}_{12}$ is $(p - q) \times q$, and $\boldsymbol{I}_{22}$ is $q \times q$. Similarly, partition $\boldsymbol{I}_1^{-1}(\boldsymbol{\beta})$ as:

$$\mathbf{I}_1^{-1}(\boldsymbol{\beta}) = \begin{pmatrix} \mathbf{I}^{11} & \mathbf{I}^{12} \\ \mathbf{I}^{21} & \mathbf{I}^{22} \end{pmatrix} \tag{9}$$

where the dimensions of $\mathbf{I}^{11}$, $\mathbf{I}^{12}$, $\mathbf{I}^{21}$, and $\mathbf{I}^{22}$ are the same as those of $\mathbf{I}_{11}$, $\mathbf{I}_{12}$, $\mathbf{I}_{12}^T$, and $\mathbf{I}_{22}$, respectively. Let $\mathbf{I}_{22.1} = \mathbf{I}_{22} - \mathbf{I}_{12}^T \mathbf{I}_{11}^{-1} \mathbf{I}_{12}$. Then $\mathbf{I}^{22} = \mathbf{I}_{22.1}^{-1}$ by the formula for the inverse of a partitioned matrix (e.g., Harville, 2008, Section 8.5). Thus, the score statistic in the test of $H_0: \boldsymbol{\beta}_2 = \mathbf{0}$ versus $H_A: \boldsymbol{\beta}_2 \neq \mathbf{0}$ is given by:

$$n^{-1} \begin{bmatrix} \mathbf{S}_1(\widehat{\boldsymbol{\beta}}) \\ \mathbf{S}_2(\widehat{\boldsymbol{\beta}}) \end{bmatrix}^T \mathbf{I}_1^{-1}(\widehat{\boldsymbol{\beta}}) \begin{bmatrix} \mathbf{S}_1(\widehat{\boldsymbol{\beta}}) \\ \mathbf{S}_2(\widehat{\boldsymbol{\beta}}) \end{bmatrix}$$

$$= n^{-1} \begin{bmatrix} \mathbf{0} \\ \mathbf{S}_2(\widehat{\boldsymbol{\beta}}) \end{bmatrix}^T \begin{bmatrix} \mathbf{I}^{11}(\widehat{\boldsymbol{\beta}}) & \mathbf{I}^{12}(\widehat{\boldsymbol{\beta}}) \\ \mathbf{I}^{21}(\widehat{\boldsymbol{\beta}}) & \mathbf{I}^{22}(\widehat{\boldsymbol{\beta}}) \end{bmatrix} \begin{bmatrix} \mathbf{0} \\ \mathbf{S}_2(\widehat{\boldsymbol{\beta}}) \end{bmatrix}$$

$$= n^{-1} [\mathbf{S}_2(\widehat{\boldsymbol{\beta}})]^T \mathbf{I}^{22}(\widehat{\boldsymbol{\beta}}) [\mathbf{S}_2(\widehat{\boldsymbol{\beta}})] . \tag{10}$$

Under the null hypothesis, $n^{-1/2} \mathbf{S}_2(\widehat{\boldsymbol{\beta}}) \xrightarrow{d} N_q(\mathbf{0}, \mathbf{I}_{22.1}[(\boldsymbol{\beta}_1^T, \mathbf{0}^T)^T])$. Thus, the asymptotic distribution of the score statistic in Equation 10 is that of a $\chi^2(q)$ random variable.

Many researchers in educational measurement have used score tests to detect model under-specification in their respective areas of interest; perhaps the best known application is in the field of structural equation modeling (SEM) where Sörbom (1989) described the use of one degree of freedom score tests that he referred to as modification indices. SEM is a broad term encompassing many related modeling families, each with the primary goal of explaining the covariance structure among a set of variables. Traditionally, the observed variables in SEM can be either categorical or continuous but all latent variables must be continuous, thereby excluding DCMs from the SEM framework. Measures of overall model fit assess whether the structural equation model hypothesized by the researcher fits the observed data adequately. If the fit is poor, modification indices can be used as a guide in determining which parameters to *add* to the model so as to significantly improve model-data fit, i.e. they test for model under-specification.

In confirmatory factor analysis, the model for the vector of observed variables $Y$ is:

$$Y = \tau + \Lambda\theta + \varepsilon \tag{11}$$

where $\tau$ is a vector of intercept parameters, $\Lambda$ is a matrix of regression weights commonly referred to as factor loadings with number of rows equal to the number of observed variables and number of columns equal to the number of latent variables, $\theta$ is a vector of the continuous latent variables referred to as factors, and $\varepsilon$ is a vector of measurement errors uncorrelated with $\theta$. As a simple example of how modification indices can be applied to the measurement component of a structural equation model, consider the following hypothesized factor loading matrix for a confirmatory factor model with five observed and two latent variables:

$$\Lambda = \begin{bmatrix} \lambda_{1,1} & 0 \\ \lambda_{2,1} & \lambda_{2,2} \\ \lambda_{3,1} & 0 \\ 0 & \lambda_{4,2} \\ \lambda_{5,1} & \lambda_{5,2} \end{bmatrix}. \tag{12}$$

The '0' entry in the first column means the fourth measured variable is not hypothesized to be an indicator of the first latent factor, and there are '0' entries in the second column because the first and third measured variables are not hypothesized as indicators of the second factor. If the model is a poor fit for the data, then allowing some of the parameters constrained to zero to be freely estimated may improve the fit. For example, adding a path from the first factor to the fourth measured variable may significantly reduce the discrepancy between model and data. The modification index for making this determination is the score statistic (see Equation 10) in a test of $H_0: \lambda_{4,1} = 0$ versus $H_A: \lambda_{4,1} \neq 0$, which will have a $\chi^2(1)$ distribution for large samples.

There are several paths that could be added and the determination of which ones should be included in the model is typically made in a sequential forward selection procedure. Such a process begins by calculating the modification index for all constrained paths and the most significant ones are added to the model one at a time until it is no longer possible to improve model fit by freely estimating one of the constrained parameters. However, as MacCallum et al. (1992) point out, in making multiple successive modifications to a model one runs the risk of capitalizing on chance variation in the sample data such that the model modifications may not generalize to the population. Furthermore, some modifications suggested by such a procedure may not have a meaningful interpretation, making it important for researchers to carefully consider the substantive implications of each potential modification.

Another important criticism concerning the typical use of score tests in the context of SEM is that users rarely control for Type I error rates across multiple tests of individual parameters, even though they likely would do so in the context of an analysis of variance (e.g., Cribbie, 2007; Green & Babyak, 1997). To address this lack of multiplicity control when modification indices are used in SEM, Green and Babyak (1997) demonstrated the use of three methods for controlling Type I error rates in both a path analytic example and a factor analytic example, including the well-known Bonferroni procedure (Dunn, 1961). These criticisms and potential resolutions also apply to modification indices developed for use with DCMs.

**Adapting score tests for DCMs**

*Q-matrix modification indices*

This method of model modification so prevalent in the SEM literature can be extended to a diagnostic modeling context, and could be used for detection of under-specification of both the Q-matrix (the focus of this subsection) and the diagnostic model (the focus of the next subsection). As an example of how modification indices would function in a test of Q-matrix under-specification, consider a hypothesized Q-matrix for the DCM of a test with five items

measuring two attributes:

$$Q = \begin{bmatrix} 1 & 0 \\ 1 & 1 \\ 1 & 0 \\ 0 & 1 \\ 1 & 1 \end{bmatrix}. \tag{13}$$

With respect to indicating which items measure which latent variables, the Q-matrix is analogous to the factor loading matrix in Equation 12. The '0' entry in the second column of the first row implies that attribute 2 is not measured by item 1. Modification indices can determine whether the addition of this path, or any path corresponding to a '0' entry in the Q-matrix, would significantly improve the fit of the model to the sample data. However, even if a modification is statistically justifiable it may not be substantively plausible, thus the item should be reviewed to determine whether measurement of this attribute is even conceivable.

Q-matrix modification indices will be a bit more complex than their SEM counterparts due to the fact that DCMs incorporate terms representing interactions between latent variables. In SEM the latent variables are typically combined in a purely additive form, such that the addition of a path from a latent factor to an observed variable implies the addition of only one model parameter. For DCMs, the addition of a path from an attribute to an item entails the addition of a main effect and one or more interaction terms. For example, consider item 1 in the Q-matrix of Equation 13 for which the LCDM item response function is given by:

$$P(Y_{e1} = 1|\boldsymbol{\alpha}_c) = \frac{\exp(\lambda_{1,0} + \lambda_{1,1,(1)}\alpha_{c1})}{1 + \exp(\lambda_{1,0} + \lambda_{1,1,(1)}\alpha_{c1})}. \tag{14}$$

If this item were specified as measuring both attributes 1 and 2 instead of only attribute 1, the fully-specified form of the LCDM function would then be:

$$P(Y_{e1} = 1|\boldsymbol{\alpha}_c) = \frac{\exp(\lambda_{1,0} + \lambda_{1,1,(1)}\alpha_{c1} + \lambda_{1,1,(2)}\alpha_{c2} + \lambda_{1,2,(1,2)}\alpha_{c1}\alpha_{c2})}{1 + \exp(\lambda_{1,0} + \lambda_{1,1,(1)}\alpha_{c1} + \lambda_{1,1,(2)}\alpha_{c2} + \lambda_{1,2,(1,2)}\alpha_{c1}\alpha_{c2})}. \tag{15}$$

Hence, using the score statistic to test the hypotheses $H_0: \boldsymbol{\beta}_2 = \boldsymbol{0}$ versus $H_A: \boldsymbol{\beta}_2 \neq \boldsymbol{0}$ where $\boldsymbol{\beta}_2^T = \left(\lambda_{1,1,(2)}, \lambda_{1,2,(1,2)}\right)$ represents an omnibus test of whether item 1 measures attribute 2 in addition to measuring attribute 1. However, the order constraints imposed upon the $\lambda$ parameters define a complicated parameter space under this alternative hypothesis. For practitioners, implementation will be much simpler if modification indices instead focused on the individual $\lambda$ parameters included in $\boldsymbol{\beta}_2$ in a one at a time sequential fashion, as is common in SEM modification indices reported from widely used statistical software packages. Conducting individual score tests has the added benefit of immediately identifying the particular parameters that differ from zero, rather than just indicating that at least one of them differs from zero. Thus, in testing whether item 1 measures attribute 2 in addition to measuring attribute 1 there will be two Q-matrix modification indices, which we define as the score statistics in the tests of the null hypotheses: (1) $H_0: \lambda_{1,1,(2)} = 0$ and (2) $H_0: \lambda_{1,2,(1,2)} = 0$.

    The alternative hypotheses for these tests are determined in part by the order constraints imposed on the $\lambda$ parameters. Recall that, in general, main effects must be greater than zero in order for mastery of an additional measured attribute to increase (rather than decrease) the chance of answering an item correctly. Thus, for testing $H_0: \lambda_{1,1,(2)} = 0$ the alternative hypothesis is $H_A: \lambda_{1,1,(2)} > 0$. Now, the second Q-matrix modification index is testing for the addition of an interaction term between attributes 1 and 2 to the model in Equation 14, which contains only an intercept term and a main effect for attribute 1. Hence, it is only necessary to require the interaction to be greater than zero, thus for testing $H_0: \lambda_{1,2,(1,2)} = 0$ the alternative hypothesis is $H_A: \lambda_{1,2,(1,2)} > 0$. Note that the score statistic given in Equation 10 applies only to two-sided tests. However, methods do exist for conducting score tests when the alternative hypothesis of interest is one-sided (e.g., Silvapulle & Silvapulle, 1995). These methods and their

application to modification indices for diagnostic classification models will be discussed at the conclusion of this section.

The number of Q-matrix modification indices associated with a given item will depend upon the number of attributes both the item and the test are specified as measuring. For instance, given a test measuring four attributes and an item specified as measuring two of these attributes, there will be eight Q-matrix modification indices that could be considered for this item: one main effect, two two-way interactions, and one three-way interaction for each of the unspecified attributes. For long tests measuring many items, the total number of Q-matrix modification indices to consider can become quite large. In the context of SEM, it has been suggested that researchers conduct a restricted search in which only indices for the modifications which could be substantively justified are considered, thereby reducing the total number of hypothesis tests (e.g., MacCallum, 1986). For items already specified as measuring multiple attributes, it would also make sense for the researcher to initially consider Q-matrix modification indices corresponding only to the main effects and the lower-order (e.g., two-way) interactions. Given this potential for large numbers of tests, it is paramount that some sort of multiplicity correction, such as the Bonferroni procedure, is used with Q-matrix modification indices.

Consider again the example of using Q-matrix modification indices to test whether item 1 in the Q-matrix of Equation 13 measures attribute 2 in addition to attribute 1. Rejection of either $H_0: \lambda_{1,1,(2)} = 0$ or $H_0: \lambda_{1,2,(1,2)} = 0$ would suggest item 1 does measure attribute 2. If item 1 is reviewed and this suggestion seems reasonable, the Q-matrix in Equation 13 should be altered so that the entry in the second column of the first row is now a '1' instead of a '0.' Now, if only $H_0: \lambda_{1,1,(2)} = 0$ is rejected, then it would make sense for the model for item 1 to be re-specified so as to include the main effect of attribute 2. But, if only $H_0: \lambda_{1,2,(1,2)} = 0$ is rejected the analyst must decide whether or not to adhere to the principle of hierarchy in statistical modeling, whereby higher-order interaction terms are included only if all corresponding lower-order terms are also included. In this case, following the principle of hierarchy would mean including both the significant interaction between attributes 1 and 2 and the non-significant main effect for attribute 2 in the re-specified model. In general, though, it is not advisable to add multiple parameters in a subsequent model re-specification, as modification indices are a comparison of the initially specified model and a model that adds just the parameter under consideration.

Q-matrix modification indices were so named because they represent the addition of model parameters that would alter the entries of the Q-matrix. However, when the hypothesized model is not a fully-specified LCDM, e.g., the model contains only main effects and no interaction terms, it is possible to modify the model parameters in such a way that the Q-matrix is not altered. Modification indices for these model parameters will be referred to as diagnostic model modification indices, and are elaborated on in the following subsection.

*Diagnostic model modification indices*

Diagnostic modeling families such as the LCDM offer modeling flexibility and a unified DCM framework, as most of the commonly used DCM variants are simply special cases of the fully-specified LCDM. However, many researchers and analysts still choose to implement a specific restricted DCM, with the most common being the DINA model. Diagnostic model modification indices can be used to determine whether freeing some of the parameters constrained by a particular DCM variant might significantly improve model fit, thereby allowing analysts to test whether sample evidence rejects the response process hypothesized by their chosen DCM.

To define diagnostic model modification indices, consider the case where the initially specified diagnostic model is the DINA model, a noncompensatory DCM hypothesizing that all measured attributes must be mastered to have a high probability of answering an item correctly. That is, the probability of responding correctly to an item can only increase by mastering *all* measured attributes and does not increase incrementally for each additional attribute mastered. Thus, the LCDM representation of the DINA model for an item *i* measuring attributes 1 and 2 is:

$$P(Y_{ei} = 1|\boldsymbol{\alpha_e}) = \frac{\exp(\lambda_{i,0} + \lambda_{i,2,(1,2)}\alpha_{e1}\alpha_{e2})}{1 + \exp(\lambda_{i,0} + \lambda_{i,2,(1,2)}\alpha_{e1}\alpha_{e2})}. \qquad (16)$$

In comparing the DINA model to the fully-specified LCDM for an item *i* measuring attributes 1 and 2 as given in Equation 2, there will be two associated diagnostic model modification indices, which we define as the score statistics in the tests of the hypotheses: (1) $H_0: \lambda_{i,1,(1)} = 0$ versus $H_A: \lambda_{i,1,(1)} > 0$ and (2) $H_0: \lambda_{i,1,(2)} = 0$ versus $H_A: \lambda_{i,1,(2)} > 0$.

These modification indices test whether the response process hypothesized by the DINA model is supported by sample evidence. If so, then neither main effect would be statistically significant, but their interaction term would be significant. If only one main effect is significant, this item might not measure the second attribute and the two-way interaction term would not be significant in a re-specified model including the significant main effect. Thus, in light of the availability of the LCDM, initially hypothesizing a DINA model is inefficient. It would be more productive in terms of number of model specifications to begin either with a fully-specified LCDM and subsequently remove non-significant parameters, or to follow the principle of hierarchy in statistical modeling and begin with a model including only main effects and possibly some lower-order interaction terms, and then test for the inclusion of higher-order interactions.

DINA model modification indices can be constructed for all items measuring multiple attributes, with the number of modification indices depending upon the number of attributes measured by the item. For example, an item measuring three attributes will have six associated DINA model modification indices, three for the omitted main effects and three for the omitted two-way interactions. For items measuring only one attribute, the LCDM representation of the DINA model contains an intercept and one main effect, and is therefore fully specified. Thus, no DINA model modification indices will be needed for single attribute items.

As diagnostic model modification indices can be applied whenever the initial model is not a fully-specified LCDM, an important application will be to the case where higher-order interaction terms were initially omitted from the model because of their computational burden. In such circumstances, diagnostic model modification indices corresponding to these omitted interaction terms would supply information about whether their exclusion is statistically justifiable. Modification indices are computationally efficient in that they can provide such information without actually estimating the omitted model parameters.

*Score tests in a constrained parameter space*

The score statistic in Equation 10 implicitly assumes a two-sided alternative hypothesis, and must be adjusted for the one-sided cases of interest in DCM modification. Silvapulle and Silvapulle (1995) presented a score test appropriate for one-sided alternatives, and Verbeke and Molenberghs (2003) demonstrated its use in the context of variance components testing in the generalized linear mixed model. Here, we demonstrate how this one-sided score statistic can be used as a modification index for DCMs.

As outlined above, the hypotheses associated with modification indices for DCMs will frequently be of the form $H_0: \beta_2 = 0$ versus $H_A: \beta_2 > 0$. When $\beta_2$ is a scalar constrained to be greater than zero, the one-sided score statistic $T_S$ based on a sample of $E$ examinees is given by:

$$T_S = \frac{[S_2(\widehat{\boldsymbol{\beta}})]^2}{E \cdot I^{22}(\widehat{\boldsymbol{\beta}})} - \inf\left\{ \left( \frac{S_2(\widehat{\boldsymbol{\beta}})}{\sqrt{E}} - b \right)^2 I^{22}(\widehat{\boldsymbol{\beta}}) \middle| b > 0 \right\} \tag{17}$$

with $T_S \sim \frac{1}{2}\chi^2(0) + \frac{1}{2}\chi^2(1)$. Note that the first term in $T_S$ is the general score statistic, which has a $\chi^2(1)$ distribution. If unconstrained estimation of $\beta_2$ would result in a negative value of $\hat{\beta}_2$, the infimum in Equation 17 is achieved when $b = 0$, resulting in $T_S = 0$. Else, the infimum in Equation 17 is zero and $T_S$ will be the value of the general score statistic, providing an intuitive argument for why the distribution of $T_S$ is a 50:50 mixture of the $\chi^2(0)$ and $\chi^2(1)$ distributions.

When testing for the addition of an interaction term to a model containing only main effects, the alternative hypothesis will be of the form $H_A: \beta_2 > -k$, where $\beta_2$ represents an interaction term and $k$ is the value of the smallest main effect. In this case, the infimum in Equation 17 is conditional on $b > -k$, and the one-sided score statistic will follow a weighted mixture of the $\chi^2(0)$ and $\chi^2(1)$ distributions with unknown weights (Silvapulle & Sen, 2005, Section 3.5). Using $\chi^2(1)$ as the reference distribution will serve as a good approximation when the sample size is large, as is frequently the case in educational testing.

In order to evaluate $T_S$, we will need to find

$$S_2(\hat{\boldsymbol{\beta}}) = \left.\frac{\partial \ell(\boldsymbol{\beta})}{\partial \beta_2}\right|_{\boldsymbol{\beta}=\hat{\boldsymbol{\beta}}} \qquad (18)$$

which is the partial derivative of the log-likelihood of the full model (i.e., the model that includes all reduced model parameters and $\beta_2$) with respect to $\beta_2$, evaluated at the maximum likelihood estimates of the reduced model when $\beta_2 = 0$. In practice, the software package Mplus can find maximum likelihood estimates of LCDM parameters (see Templin & Hoffman, 2013). From Equation 1, we see that the log-likelihood of the LCDM for a sample of $E$ examinees is:

$$\ell = \sum_{e=1}^{E} \log f(Y_e; \boldsymbol{\beta}) = \sum_{e=1}^{E} \log\left\{\sum_{c=1}^{C} v_c \prod_{i=1}^{I} \pi_{ic}^{y_{ei}}(1-\pi_{ic})^{1-y_{ei}}\right\} \qquad (19)$$

where the $\pi_{ic}$ parameters are as defined by the LCDM of Equation 5 and the $\nu_c$ are defined by the chosen structural model. Evaluation of $S_2(\widehat{\boldsymbol{\beta}})$ requires the partial derivatives of the LCDM log-likelihood with respect to the item parameters, which can be shown to be:

$$\frac{\partial \ell}{\partial \lambda_{i,l,(\boldsymbol{a})}} = \sum_{e=1}^{E} \frac{\sum_{c=1}^{C} \nu_c \frac{\partial \eta_i}{\partial \lambda_{i,l,(\boldsymbol{a})}} (y_{ei} - \pi_{ic}) \prod_{i=1}^{I} \pi_{ic}^{y_{ei}} (1 - \pi_{ic})^{1-y_{ei}}}{\sum_{c=1}^{C} \nu_c \prod_{i=1}^{I} \pi_{ic}^{y_{ei}} (1 - \pi_{ic})^{1-y_{ei}}} \quad (20)$$

where $\eta_i = \lambda_{i,0} + \boldsymbol{\lambda}_i^T \boldsymbol{h}(\boldsymbol{\alpha}_c, \boldsymbol{q}_i)$ as defined in Equation 5 and the entries of $\boldsymbol{\lambda}_i$ are denoted as $\lambda_{i,l,(\boldsymbol{a})}$ where $l$ designates the level of the effect and the vector $\boldsymbol{a}$ identifies the attributes with which the parameter is associated.

The calculation of $I^{22}(\widehat{\boldsymbol{\beta}})$ is also needed to evaluate $T_S$, which depends upon the entries in the information matrix $\mathbf{I}_1(\boldsymbol{\beta})$, defined in Equation 8. However, as discussed in McLachlan and Peel (2000, Section 2.15), calculation of the second-order derivatives of the model log-likelihood can be quite tedious for mixture models such as the LCDM. They suggest approximating the sample information matrix, i.e., $E \cdot \mathbf{I}_1(\boldsymbol{\beta})$, using the empirical observed information matrix:

$$\sum_{e=1}^{E} \frac{\partial \log f(Y_e; \widehat{\boldsymbol{\beta}})}{\partial \boldsymbol{\beta}} \times \frac{\partial \log f(Y_e; \widehat{\boldsymbol{\beta}})}{\partial \boldsymbol{\beta}^T}. \quad (21)$$

In addition to the partial derivatives with respect to the item parameters, this approximation also requires the partial derivatives of the LCDM log-likelihood with respect to the structural parameters. For the log-linear structural model proposed in Henson & Templin (2005), the expected number of examinees in a latent class, $\mu_c$, is predicted by:

$$\log(\mu_c) = \sum_{a=1}^{A} \gamma_{1,(a)} \alpha_{ca} + \sum_{a=1}^{A-1} \sum_{a'=a+1}^{A} \gamma_{2,(a,a')} \alpha_{ca} \alpha_{ca'} + \cdots + \gamma_{A,(a,a',\ldots)} \prod_{a=1}^{A} \alpha_{ca} \quad (22)$$

and the partial derivatives of the LCDM log-likelihood with respect to the structural parameters $\gamma_{l,(a)}$, where $l$ designates the level of the effect and the vector $\boldsymbol{a}$ identifies the attributes with which the structural parameter is associated, can be shown to be:

$$\frac{\partial \ell}{\partial \gamma_{l.(a)}} = \sum_{e=1}^{E} \frac{\sum_{c=1}^{C} \nu_c \left[ \frac{\partial}{\partial \gamma_{l.(a)}} \mu_c - \frac{\sum_{c=1}^{C} \frac{\partial}{\partial \gamma_{l.(a)}} \mu_c}{\sum_{c=1}^{C} \mu_c} \right] \left[ \prod_{i=1}^{I} \pi_{ic}^{y_{ei}} (1-\pi_{ic})^{1-y_{ei}} \right]}{\sum_{c=1}^{C} \nu_c \prod_{i=1}^{I} \pi_{ic}^{y_{ei}} (1-\pi_{ic})^{1-y_{ei}}}. \quad (23)$$

Mplus can provide the approximation to the sample information matrix defined in Equation 21 for the estimated model if MLF is specified as the estimator in the analysis statement, where MLF requests maximum likelihood parameter estimates with standard errors approximated by first-order derivatives (Muthén & Muthén, 1998-2017). The approximation in Equation 21 is based on work by Louis (1982), and was also used by Glas (1999) and by Glas and Suárez-Falcón (2003) in their application of the score test to item response theory models.

**Simulation study**

To assess the utility of both Q-matrix and diagnostic model modification indices as methods for detecting under-specification of DCMs, a simulation study was conducted. The simulation study consisted of two main components: a study evaluating the performance of Q-matrix modification indices and a study evaluating diagnostic model modification indices. Each of these included both a Type I error study and a power analysis. In the Type I error studies, the goal was to demonstrate that modification indices for DCMs indicate that unnecessary attributes or model parameters should be added to the model at an acceptably low rate (i.e., at the Type I error rate specified by the researcher). In the power analyses, the goal was to verify that modification indices for DCMs indicate that necessary attributes or model parameters should be added to the model at an acceptably high rate (i.e., that the test is powerful), and to investigate the sample sizes needed to reach a desired level of power. The simulation study evaluating Q-matrix modification indices also included a Q-matrix recovery study in which the goal was to evaluate the performance of Q-matrix modification indices in the presence of both under-specification and over-specification of the Q-matrix, as commonly occurs in practice.

All simulation conditions included 30 items, 3 attributes, a .455 tetrachoric correlation among attributes, and 1000 replications. In the Type I error studies and power analyses, samples of 500, 1000, 2500, and 5000 examinees were considered. The Q-matrix recovery study included samples of 1000 and 2500 examinees. In all cases, the Q-matrix for the generating models was balanced, with every item measuring either one or two attributes and each possible pattern (100, 010, 001, 110, 101, and 011) repeated five times. The item parameter values of the generating models were selected based upon the resulting item response probabilities. The item intercepts were set to $-1.5$ in all cases, such that examinees having mastered none of the measured attributes respond correctly with probability .18, roughly equivalent to the chance of guessing the correct answer from among five answer choices. For the item main effects and interaction terms, two different cases were considered: a smaller effect size in which examinees that have mastered all measured attributes respond correctly with probability .62 and a larger effect size in which this probability is .92. Thus, items in the larger effect size will be better able to discriminate between masters and non-masters of the measured attributes. Altogether, the Type I error studies and power analyses both included eight conditions (4 sample sizes × 2 effect sizes) in each of the two modification index studies (Q-matrix and diagnostic model) and the Q-matrix recovery study included four conditions (2 sample sizes × 2 effect sizes), for a total of 36 distinct conditions.

The number of items, attributes, and examinees chosen for this simulation study are reflective of values currently used in practice and in other studies reported in the DCM literature. A test with 30 items would be typical of the length of a formative assessment such as a county-wide benchmark test. Setting the number of attributes to three would ensure that each attribute could be measured a sufficient number of times by the test. A sample of 2500 examinees would be representative of the number of students an average sized county would have per grade level for their benchmark testing program. Moreover, to investigate likelihood-based item-level fit statistics in DCMs, the simulation study of Ma et al. (2016) used 30 items, 5 attributes, and samples of 500, 1000, and 2000 examinees, then Sorrel et al. (2017) used tests with 12, 24, and 26 items measuring 4 attributes and samples of 500 and 1000 examinees. To investigate methods for correcting Q-matrix misspecification, the simulation study of Kunina-Habenicht et al. (2012) contained tests with 25 and 50 items measuring both three and five attributes, and samples of 1000 and 10000 examinees, then Liu et al. (2012) used a Q-matrix with 20 items and three attributes, with sample sizes ranging from 500 to 4000 examinees.

All models in the simulation studies were estimated in Mplus, and the modification indices were calculated using a program written in the statistical software package R. Item parameter recovery was assessed by comparing the generating values to the mean estimated values across the 1000 replications in a particular simulation condition, and the largest observed discrepancy was 0.01.

 *Type I error study for Q-matrix modification indices*

This study considered modification indices for the addition of attribute 2 to the model for item 1, which measures only attribute 1 in the generating model. That is, the unnecessary addition of both $\lambda_{1,1,(2)}$ and $\lambda_{1,2,(1,2)}$ to the estimated LCDM was considered. The proportion of replications in which the modification indices incorrectly indicated this modification was advantageous is summarized in Table 1. These observed Type I error rates were consistently close to the nominal significance level $\alpha$ across the range of reported $\alpha$ values (.10, .05, .025, .01, and .005). Given that two hypothesis tests were conducted on each set of simulated data, one for the main effect and one for the interaction term, the familywise error rate is of concern. For example, with 2500 examinees the observed probability that at least one of the two null hypotheses was incorrectly rejected at the $\alpha = .05$ level was .107 in the large effect size and .069 in the smaller effect size. Hence, some type of multiplicity control should be considered when using Q-matrix modification indices in practice.

[Insert Table 1 about here]

*Power analysis for the Q-matrix modification indices*

The item of interest in this study was item 4, which measures both attributes 1 and 2 in the generating Q-matrix, but was incorrectly specified as only measuring attribute 1 in the estimated model (i.e., $\lambda_{4,1,(2)}$ and $\lambda_{4,2,(1,2)}$ were both included in the generating but not the estimated LCDM). The proportion of replications in which the modification indices correctly detected this under-specification are given in Table 2. The significance levels reported include: (1) $\alpha = .05$, corresponding to no multiple testing correction, (2) $\alpha = .025$, corresponding to a Bonferroni correction for the two tests actually conducted for each sample, and (3) $\alpha = .0005$, corresponding to a Bonferroni correction for the 105 potential tests for the main effect and two-way interaction of every '0' entry in the generating Q-matrix. As seen in Table 2, these tests were quite powerful even when the familywise error rate was controlled for the 105 potential tests.

[Insert Table 2 about here]

*Q-matrix recovery study*

In practice, Q-matrices can contain both under-specification and over-specification such that Q-matrix modification indices would need to be used in tandem with a method for detecting over-specification, such as the Wald test reported in Mplus, in order to recover the true generating Q-matrix. To evaluate how Q-matrix modification indices perform in such situations, the estimated model in the Q-matrix recovery simulation study both over-specified item 1 as measuring attributes 1 and 2, when it measures only attribute 1 in the generating model, and under-specified item 4 as measuring only attribute 1, when it measures both attributes 1 and 2 in the generating model. In practice, we would recommend that analysts begin their model refinement by first using the Wald test results reported in Mplus to guide their decisions as to removing any non-significant item parameters before using Q-matrix modification indices to suggest significant item parameters to potentially add to the model. Following this approach, and adhering to the principle of hierarchy in statistical modeling, the estimated model in the Q-matrix recovery simulation study would require three model refinement steps to recover the generating Q-matrix.

- Step 1: The Wald $Z$-statistic for $\lambda_{1,2,(1,2)}$ is not statistically significant and this parameter is removed from the model. In this case, the null hypothesis does not place the parameter at a boundary value and statistical significance can be assessed according to the standard normal distribution.
- Step 2: In the estimated model without the interaction term for item 1, the Wald $Z$-statistic for $\lambda_{1,1,(2)}$ is not statistically significant and this parameter is removed from the model. As the null hypothesis for this test places the parameter at its lower bound of 0, statistical significance of $Z^2$ should be assessed according to the $\frac{1}{2}\chi^2(0) + \frac{1}{2}\chi^2(1)$ distribution (Molenberghs & Verbeke, 2007). Note that removing both

$\lambda_{1,2,(1,2)}$ and $\lambda_{1,1,(2)}$ from the model changes the Q-matrix entry for attribute 2 of item 1 from a '1' to a '0'.

- Step 3: In the estimated model with item 1 correctly specified as measuring only attribute 1, the modification indices for $\lambda_{4,1,(2)}$ and/or $\lambda_{4,2,(1,2)}$ are statistically significant, correctly suggesting to analysts that Item 4 also measures attribute 2. As previously discussed, statistical significance of these Q-matrix modification indices should be assessed according to the $\frac{1}{2}\chi^2(0) + \frac{1}{2}\chi^2(1)$ distribution.

This model refinement process was followed in every replication of the Q-matrix recovery simulation study, and the proportion of correct decisions at each step with a significance level of $\alpha = .05$ is given in Table 3. Additionally, the last column of Table 3 gives the proportion of replications in which a correct decision was made at every step such that the generating Q-matrix was recovered by this model refinement process.

[Insert Table 3 about here]

The Q-matrix recovery rate was quite high across all simulation conditions with $\alpha = .05$. In fact, the proportion of correct decisions in Steps 1 and 2 was higher than the expected rate of $1 - \alpha$. If a Bonferroni correction is applied and the significance level is adjusted for the number of tests of interest at each step (60 in Step 1, 59 in Step2, and 105 in Step 3), the proportion of correct decisions in Step 1 becomes 1.000 in all conditions as it already was for Step 2 when $\alpha = .05$. The results for Step 3 are in line with the results from the power analysis when $\alpha = .0005$, where the Q-matrix modification indices were not statistically significant in just a few replications for the 1000 examinee smaller effect size condition, including 2 replications where neither $\lambda_{4,1,(2)}$ nor $\lambda_{4,2,(1,2)}$ were statistically significant such that the generating Q-matrix would not be recovered.

*Type I error study for diagnostic model modification indices*

This study focused on modification indices for the DINA model because of its popularity among researchers and analysts. The DINA model was both the generating and estimated model for all items in this study in order to estimate the ability of diagnostic model modification indices to correctly find that a given diagnostic model is *not* under-specified. Specifically, we considered modification indices for the main effects of attributes 1 and 2 in the model for item 4, which were not in the generating model. Table 4 reveals that the observed Type I error rates were consistently close to the nominal level across the range of reported $\alpha$ values (.10, .05, .025, .01, and .005). Observed familywise error rates with no multiplicity control were again inflated, with the observed probability that at least one of the two null hypotheses was incorrectly rejected at the $\alpha$ = .05 level was .089 in the large effect size and .081 in the smaller effect size with 2500 examinees.

[Insert Table 4 about here]

*Power analysis for diagnostic model modification indices*

In this study, data were generated from a fully-specified LCDM for the given Q-matrix. However, the DINA model parameters were estimated for item 4 such that the main effects for attributes 1 and 2 were included in the generating but not in the estimated model. As seen in Table 5, the modification indices were quite powerful in the detection of this under-specification for the large effect size conditions. However, they were less powerful for the smaller effect sizes where the items were not as discriminating between masters and non-masters of the measured attributes, especially for smaller sample sizes and when using a significance level of $\alpha$ = .0017 to control the familywise error rate for the 30 tests that would result if the DINA was specified for all items.

[Insert Table 5 about here]

**DTMR fractions test data analysis**

Having defined a one-sided score statistic appropriate for use as a modification index for DCMs, affirmed Type I error control when a mixture $\chi^2$ reference distribution is used, and explored the conditions in which these modification indices have reasonable power, we next investigated their utility to suggest appropriate model revisions in practice. The data used in this analysis were from a large-scale administration of the Diagnosing Teachers' Multiplicative Reasoning (DTMR) Fractions Test, a diagnostic test designed to assess middle grades teachers' conceptual understandings of fraction arithmetic (Bradshaw et al., 2014). The DTMR Factions Test was specifically designed for assessing examinee mastery of multiple attributes using DCMs, in contrast to typical analyses where DCMs are fit to existing response data from exams developed for use with other (often unidimensional) psychometric models.

The DTMR fractions test included 21 question stems and 28 items in total. The test was designed to measure four essential attributes of multiplicative reasoning: attending to referent units ($\alpha_1$), partitioning and iterating ($\alpha_2$), identifying appropriate situations to make multiplicative comparisons ($\alpha_3$), and forming multiplicative comparisons ($\alpha_4$). The test was administered to a sample of 990 in-service middle-grades mathematics teachers from across the country. Bradshaw et al. (2014) analyzed the response data using a fully-specified LCDM with the initially hypothesized Q-matrix given in Table 6. Note that there is not an entry for item 20 because this item was removed from the analysis due to difficulties in scoring the responses. Item parameters removed from the model on the basis of Wald test results led to the seven changes in the Q-matrix noted in Table 6.

[Insert Table 6 about here]

*Q-Matrix modification indices for the DTMR fractions test data*

Q-matrix modification indices were used to test for under-specification of the Q-Matrix using the estimated model of Bradshaw et al. (2014). That is, for each '0' entry in the initial Q-matrix in Table 6, a Q-matrix modification index was calculated to determine if there is statistical evidence that the item measures that attribute. In an effort to reduce the total number of hypothesis tests, only modification indices corresponding to main effects and two-way interactions of items specified as measuring multiple attributes were considered. The results are given in Table 7.

[Insert Table 7 about here]

As there were 148 potential model modifications considered in Table 7, a Bonferroni correction to control the familywise error rate at .05 required $p < (.05/148)$ for statistical significance. This corresponded to a critical value of 11.55 for the $\frac{1}{2}\chi^2(0) + \frac{1}{2}\chi^2(1)$ reference distribution. From Table 7, we see that 10 modification indices exceeded 11.55. There were four statistically significant two-way interaction terms, and in each case the corresponding main effect was also significant. Hence, the Q-matrix modification indices suggested six possible alterations to the initial Q-matrix: specifying Item 2 as also measuring $\alpha_2$, specifying item 3 as also measuring $\alpha_1$, specifying Item 6 as also measuring $\alpha_1$ and $\alpha_4$, and specifying item 8d as also measuring $\alpha_1$ and $\alpha_2$. This represents a reasonable number of Q-matrix modifications for the mathematics education content specialists to consider as a means to improving the agreement between the statistical model and the observed response patterns. Bradshaw (2017) noted that item 3 was a difficult item with only 40% of examinees having mastered $\alpha_2$, the only attribute the item was originally specified as measuring, expected to answer the item correctly and that an additional required attribute may explain the difficulty of this item.

*DINA model modification indices for the DTMR fractions test data*

In the second component of the DTMR Fractions test data analysis, DINA model modification indices were used to determine if this popular DCM might be an appropriate model for the DTMR data. At first, the parameters of the DINA model according to the initial Q-matrix in Table 6 were estimated. However, none of the items were initially specified as measuring only $\alpha_4$. This caused the attribute profile for masters of only attribute 4 to be indistinguishable from the attribute profile for masters of none of the attributes due to the DINA model parameterization (Madison & Bradshaw, 2015; Rupp & Templin, 2008). To resolve this issue, item 10a was specified as measuring only attribute 4 in a subsequent estimation of the DINA model parameters, as the LCDM analysis indicated that item 10a did not also measure attribute 1 as initially hypothesized.

As noted previously, the DINA model is equivalent to the LCDM in the case of items measuring just one attribute. For the 13 items now specified as measuring two attributes, diagnostic model modification indices were used to determine if inclusion of an omitted main effect would significantly improve model-data fit. These results are given in Table 8. If a Bonferroni correction is used to control the familywise error rate at .05 for these 26 tests, then a modification index greater than 8.36 is considered statistically significant. From Table 8, we see that there were three significant DINA model modification indices and that the suggested model modifications include: adding the main effect for attribute 3 to the model for item 8a, adding the main effect for attribute 4 to item 10b, and adding the main effect for attribute 4 to the model for item 10c. However, average estimated item parameter values were similar to those in the smaller effect size conditions of the simulation study, where estimated power for 1,000 examinees was very low (about .10) when controlling the familywise error rate at .05 for 30 tests (see Table 5).

[Insert Table 8 about here]

Several of the results from the DINA model modification indices for the DTMR data support the argument that initially hypothesizing a DINA model is inefficient. For example, consider the model specification for item 8a. The original Q-matrix identified this item as measuring attributes 3 and 4. In the LCDM analysis of Bradshaw et al. (2014), neither the interaction between these two attributes nor the main effect of attribute 4 were statistically significant and the item was subsequently re-specified as measuring only attribute 3. However, in the DINA model analysis, arriving at the same conclusion took an additional step. In the first specification, the interaction between attributes 3 and 4 was statistically significant, but the DINA model modification indices indicated that the main effect for attribute 3 should be added to the model. When the model was re-specified accordingly, the main effect for attribute 3 was significant ($z = 7.68$, $p < .001$) but the interaction term was no longer statistically significant ($z = 0$, $p = .50$). Thus, it took a third model specification to arrive at the same conclusion the LCDM analysis arrived at in two steps: Item 8a only measures attribute 3. Therefore, an analysis following the principle of hierarchy in statistical modeling by beginning with a fully-specified LCDM and subsequently removing non-significant parameters would be the preferred approach.

**Discussion**

The primary aim in applying DCMs to the analysis of item response data is to classify examinees according to their mastery of multiple latent attributes. However, misspecification in either the parameterization of the DCM or its associated Q-matrix (or both) can cause the accuracy with which examinees are classified to the correct mastery profile to diminish. There are currently limited avenues for identifying such sources of misfit which can be feasibly implemented. The modification indices for DCMs defined in this paper represent a computationally efficient inference-based method for evaluating the appropriateness of a diagnostic model specification at the item level and determining if its Q-matrix is complete. They also have the advantage of being a familiar model refinement technique in the educational measurement community because of their widespread use in SEM. The *diagnostic model modification indices* we defined apply when a reduced DCM is fit and the addition of item parameters from a more general DCM that would not alter the Q-matrix entries is being considered, whereas *Q-matrix modification indices* are used to test for the addition of item parameters that would alter the Q-matrix entries. The simulation study we conducted made important strides in understanding the conditions in which modification indices for DCMs will be most useful.

In practice, we recommend initially estimating a fully-specified LCDM and first using Wald statistics to identify non-significant item parameters to consider for removal. Note that some of the suggested modifications would change certain Q-matrix entries from a '1' to a '0' and some would not alter the Q-matrix, such that this process tests for both model and Q-matrix over-specification. Next, diagnostic model modification indices can be used to address potential model under-specification. If there are items measuring many attributes, then some higher-order interaction terms could be omitted in the initial specification, as they can be computationally intensive to estimate, and diagnostic model modification indices can be used to determine if their omission is statistically justifiable. Though initial specification of a DINA model proved inefficient in the DTMR data analysis, practitioners choosing to initially fit a reduced DCM such as the DINA model could use diagnostic model modification indices to justify their choice. The DINA model modification indices in the simulation study did have limited power in the small effect size conditions, and we believe this is because the missing main effects were quite small and as a result the intercept term served to hide those effects overall. For example, in the generating LCDM for the smaller effect size, the probability of a correct response was .18 for masters of neither measured attribute, .32 for masters of only one of the two attributes, and .62 for masters of both attributes. Across the 1000 replications of the 2500 examinee condition where the DINA model was incorrectly estimated, the mean probability of a correct response was .63 for masters of both attributes, but was .25, the average of .18 and .32, for masters of only one and for masters of none of the attributes. Sorrel et al. (2017) also found smaller item parameter sizes to adversely impact power of the score test in the diagnostic modeling context, especially with small samples. This underscores the need for items to be highly discriminating between masters and non-masters of the target attributes, which can be somewhat of an art form for item developers. After addressing model misspecification, Q-matrix modification indices can be used to detect Q-matrix under-specification. The results of the simulation study showed Q-matrix modification indices to be very powerful in the detection of an incomplete Q-matrix. The DTMR Fractions test data analysis illustrated this recommended process and how the incorporation of modification indices for DCMs into an analysis of diagnostic testing data can be useful in practice.

  Though the conditions considered in our simulation study are in no way exhaustive, they were carefully chosen so as to be reflective of those encountered in practice such that it is reasonable to assume these findings will be fairly generalizable to empirical applications. In the simulation study, we fixed the number of attributes to three in all conditions, but in practice, tests can be designed to measure more than three attributes, such as the DTMR Fractions test which measures four attributes. However, even when a test is designed to measure a large number of attributes, each item typically only measures one or two attributes, just like in our simulation conditions, as writing complex items measuring multiple attributes at once can be difficult. Though, the number of Q-matrix modification indices to consider will increase as the number of attributes measured by the test increases. For example, in the DTMR Fractions test data analysis, we chose to restrict the Q-matrix modification indices considered to only main effects and two-way interaction terms for every '0' entry in the Q-matrix. This meant that for every item specified as measuring only one attribute, there was one main effect and one two-way interaction term to consider adding for each of the three '0' Q-matrix entries for the item, for a total of six modification indices. For these items, two more modification indices would be considered for each additional attribute the test was designed to measure. Similarly, for items specified as measuring two attributes, there was one main effect and two two-way interactions to consider for both of the '0' Q-matrix entries for the item, for a total of six modification indices. For these items, three more modification indices would be considered for each additional attribute the test was designed to measure. Thus, the three-attribute simulation study may not have captured what would happen with, say, a 10-attribute test, even if each item still only measures one or two attributes. In structural equation modeling, many of the popular software packages handle this issue of large numbers of modification indices to consider by only reporting those that are larger than a specified threshold value.

 It is hoped that through the development of modification indices for DCMs and the evaluation of their statistical properties, educational researchers will have a valuable set of likelihood-based inferential procedures that can be used to justify their choice of model at the item level, to modify it as appropriate, and to take full advantage of the flexibility afforded by the LCDM family of models. This aligns with what Jöreskog (1993) referred to as a model generating approach. An additional illustration of how modification indices for DCMs fit in to this iterative process of model refinement is described in Bradshaw (2017). Thus, with the advent of diagnostic modeling families and the development of modification indices for DCMs, diagnostic model building will be able to employ empirically driven methods to arrive at a model that is substantively meaningful, reasonably parsimonious, and statistically well-fitting.

Yu, X. & Cheng, Y. (2020). Data-driven Q-matrix validation using a residual-based statistic in cognitive diagnostic assessment. *British Journal of Mathematical and Statistical Psychology*, *73*, 145-179. https://doi.org/10.1111/bmsp.12191

Table 1. Observed Type I error rates in the Q-Matrix MIs simulation study

| Effect Size | MI | Sample Size | α Level | | | | |
|---|---|---|---|---|---|---|---|
| | | | .100 | .050 | .025 | .010 | .005 |
| Large | $\lambda_{1,1,(2)}$ | 500 | .115 | .061 | .032 | .012 | .007 |
| | | 1,000 | .109 | .064 | .034 | .017 | .006 |
| | | 2,500 | .103 | .055 | .027 | .008 | .003 |
| | | 5,000 | .093 | .047 | .029 | .016 | .011 |
| | $\lambda_{1,2,(1,2)}$ | 500 | .106 | .065 | .035 | .017 | .004 |
| | | 1,000 | .091 | .050 | .026 | .009 | .005 |
| | | 2,500 | .117 | .067 | .031 | .008 | .005 |
| | | 5,000 | .097 | .046 | .022 | .008 | .003 |
| Smaller | $\lambda_{1,1,(2)}$ | 500 | .125 | .060 | .033 | .018 | .010 |
| | | 1,000 | .106 | .061 | .038 | .013 | .004 |
| | | 2,500 | .082 | .048 | .025 | .010 | .005 |
| | | 5,000 | .102 | .045 | .024 | .008 | .003 |
| | $\lambda_{1,2,(1,2)}$ | 500 | .119 | .061 | .026 | .017 | .014 |
| | | 1,000 | .117 | .058 | .027 | .012 | .007 |
| | | 2,500 | .090 | .045 | .026 | .010 | .006 |
| | | 5,000 | .101 | .052 | .019 | .007 | .006 |

*Note*: MI = modification index; α = significance level. Type I error rate calculated as the proportion of observed MIs for a given item parameter exceeding the upper α critical value of the $\frac{1}{2}\chi^2(0) + \frac{1}{2}\chi^2(1)$ distribution, where the critical value $c$ is such that $\frac{1}{2}P(\chi^2(1) > c) = \alpha$.

Table 2. Proportion of significant MIs in the Q-matrix MIs power analysis

| Effect Size | MI | Sample Size | α Level .05 | .025 | .0005 |
|---|---|---|---|---|---|
| Large | $\lambda_{4,1,(2)}$ | 500 | 1.000 | 1.000 | 1.000 |
| | | 1,000 | 1.000 | 1.000 | 1.000 |
| | | 2,500 | 1.000 | 1.000 | 1.000 |
| | | 5,000 | 1.000 | 1.000 | 1.000 |
| | $\lambda_{4,2,(1,2)}$ | 500 | 1.000 | 1.000 | 1.000 |
| | | 1,000 | 1.000 | 1.000 | 1.000 |
| | | 2,500 | 1.000 | 1.000 | 1.000 |
| | | 5,000 | 1.000 | 1.000 | 1.000 |
| Smaller | $\lambda_{4,1,(2)}$ | 500 | .996 | .989 | .858 |
| | | 1,000 | 1.000 | 1.000 | .992 |
| | | 2,500 | 1.000 | 1.000 | 1.000 |
| | | 5,000 | 1.000 | 1.000 | 1.000 |
| | $\lambda_{4,2,(1,2)}$ | 500 | .991 | .978 | .790 |
| | | 1,000 | 1.000 | 1.000 | .991 |
| | | 2,500 | 1.000 | 1.000 | 1.000 |
| | | 5,000 | 1.000 | 1.000 | 1.000 |

*Note*: MI = modification index; α = significance level. Statistical significance assessed according to the $\frac{1}{2}\chi^2(0) + \frac{1}{2}\chi^2(1)$ reference distribution.

Table 3. Proportion of correct decisions at each step of the Q-matrix recovery study with $\alpha = .05$

| | | Model Refinement Step | | | |
|---|---|---|---|---|---|
| Effect Size | Sample Size | (1) Remove $\lambda_{1,2,(1,2)}$ | (2) Remove $\lambda_{1,1,(2)}$ | (3) Add $\lambda_{4,1,(2)}$ and/or $\lambda_{4,2,(1,2)}$ | Recovered Q-matrices |
| Large | 1,000 | .976 | 1.000 | 1.000 | .976 |
|  | 2,500 | .986 | 1.000 | 1.000 | .986 |
| Smaller | 1,000 | .994 | 1.000 | 1.000 | .994 |
|  | 2,500 | .994 | 1.000 | 1.000 | .994 |

Table 4. Observed Type I error rates in the diagnostic model MIs simulation study

| Effect Size | MI | Sample Size | α Level | | | | |
|---|---|---|---|---|---|---|---|
| | | | .100 | .050 | .025 | .010 | .005 |
| Large | $\lambda_{4,1,(1)}$ | 500 | .099 | .050 | .023 | .005 | .004 |
| | | 1,000 | .090 | .045 | .026 | .011 | .004 |
| | | 2,500 | .089 | .048 | .022 | .007 | .002 |
| | | 5,000 | .087 | .044 | .022 | .010 | .005 |
| | $\lambda_{4,1,(2)}$ | 500 | .098 | .052 | .020 | .006 | .001 |
| | | 1,000 | .099 | .049 | .020 | .008 | .008 |
| | | 2,500 | .096 | .041 | .020 | .010 | .004 |
| | | 5,000 | .101 | .058 | .025 | .005 | .002 |
| Smaller | $\lambda_{4,1,(1)}$ | 500 | .097 | .043 | .024 | .007 | .000 |
| | | 1,000 | .091 | .051 | .025 | .008 | .005 |
| | | 2,500 | .090 | .041 | .018 | .009 | .003 |
| | | 5,000 | .108 | .058 | .032 | .011 | .006 |
| | $\lambda_{4,1,(2)}$ | 500 | .091 | .046 | .022 | .013 | .009 |
| | | 1,000 | .088 | .041 | .021 | .008 | .005 |
| | | 2,500 | .088 | .040 | .021 | .010 | .006 |
| | | 5,000 | .104 | .053 | .028 | .012 | .006 |

*Note*: MI = modification index; α = significance level. Type I error rate calculated as the proportion of observed MIs for a given item parameter exceeding the upper α critical value of the $\frac{1}{2}\chi^2(0) + \frac{1}{2}\chi^2(1)$ distribution, where the critical value $c$ is such that $\frac{1}{2}P(\chi^2(1) > c) = \alpha$.

Table 5. Proportion of significant MIs in the diagnostic model MIs power analysis

| Effect Size | MI | Sample Size | $\alpha$ Level | | |
|---|---|---|---|---|---|
| | | | .05 | .025 | .0017 |
| Large | $\lambda_{4,1,(1)}$ | 500 | .965 | .938 | .753 |
| | | 1,000 | 1.000 | .999 | .963 |
| | | 2,500 | 1.000 | 1.000 | 1.000 |
| | | 5,000 | 1.000 | 1.000 | 1.000 |
| | $\lambda_{4,1,(2)}$ | 500 | .952 | .930 | .744 |
| | | 1,000 | 1.000 | 1.000 | .976 |
| | | 2,500 | 1.000 | 1.000 | 1.000 |
| | | 5,000 | 1.000 | 1.000 | 1.000 |
| Smaller | $\lambda_{4,1,(1)}$ | 500 | .342 | .238 | .049 |
| | | 1,000 | .499 | .375 | .105 |
| | | 2,500 | .815 | .713 | .331 |
| | | 5,000 | .971 | .946 | .729 |
| | $\lambda_{4,1,(2)}$ | 500 | .338 | .239 | .053 |
| | | 1,000 | .478 | .360 | .105 |
| | | 2,500 | .805 | .699 | .342 |
| | | 5,000 | .964 | .939 | .729 |

*Note*: MI = modification index; $\alpha$ = significance level. Statistical significance assessed according to the $\frac{1}{2}\chi^2(0) + \frac{1}{2}\chi^2(1)$ reference distribution.

Table 6. Initial Q-matrix for the DTMR fractions test

| Item | $\alpha_1$ | $\alpha_2$ | $\alpha_3$ | $\alpha_4$ |
|---|---|---|---|---|
| 1 | 1 | 0 | 0 | 0 |
| 2 | 0 | 0 | 1 | 0 |
| 3 | 0 | 1 | 0 | 0 |
| 4 | 1 | 0 | 0 | 0 |
| 5 | 1 | 0 | 0 | $1^a$ |
| 6 | 0 | 1 | 0 | 0 |
| 7 | 1 | 0 | 0 | 0 |
| 8a | 0 | 0 | 1 | $1^a$ |
| 8b | 0 | 0 | 1 | 0 |
| 8c | 0 | 0 | 1 | 0 |
| 8d | 0 | 0 | 1 | 0 |
| 9 | 1 | 0 | 0 | 0 |
| 10a | $1^a$ | 0 | 0 | 1 |
| 10b | 1 | 0 | 0 | 1 |
| 10c | 1 | 0 | 0 | 1 |
| 11 | 1 | 0 | 0 | $1^a$ |
| 12 | 1 | 0 | 0 | 0 |
| 13 | 0 | 1 | 0 | 1 |
| 14 | 1 | 1 | 0 | 0 |
| 15a | 0 | 1 | 0 | 1 |
| 15b | 0 | 1 | 0 | $1^a$ |
| 15c | 0 | 1 | 0 | $1^a$ |
| 16 | 1 | 0 | 0 | 0 |
| 17 | 1 | 1 | 0 | 0 |
| 18 | 1 | 1 | 0 | 0 |
| 19 | 0 | 0 | $1^a$ | 0 |
| 21 | 1 | 0 | 0 | 0 |
| 22 | 1 | 1 | 0 | 0 |

*Note.* DTMR = Diagnosing Teachers' Multiplicative Reasoning; $\alpha_1$ = attending to referent units; $\alpha_2$ = partitioning and iterating; $\alpha_3$ = identifying appropriate situations to make multiplicative comparisons; $\alpha_4$ = forming multiplicative comparisons. The Q-Matrix is adapted from Bradshaw et al. (2014).
[a] Entry subsequently changed to 0 based on the statistical significance of LCDM item parameters.

Table 7. Q-Matrix modification indices for the DTMR fractions test data

| Parameter | MI | Parameter | MI | Parameter | MI | Parameter | MI |
|---|---|---|---|---|---|---|---|
| $\lambda_{1,1,(2)}$ | 8.36* | $\lambda_{7,2,(1,3)}$ | 3.48* | $\lambda_{10b,2,(2,4)}$ | 0.00 | $\lambda_{15a,2,(2,4)}$ | 0.60 |
| $\lambda_{1,2,(1,2)}$ | 0.00 | $\lambda_{7,1,(4)}$ | 0.00 | $\lambda_{10b,1,(3)}$ | 0.00 | $\lambda_{15b,1,(1)}$ | 0.00 |
| $\lambda_{1,1,(3)}$ | 11.24* | $\lambda_{7,2,(1,4)}$ | 1.84 | $\lambda_{10b,2,(1,3)}$ | 0.47 | $\lambda_{15b,2,(1,2)}$ | 0.00 |
| $\lambda_{1,2,(1,3)}$ | 0.00 | $\lambda_{8a,1,(1)}$ | 0.00 | $\lambda_{10b,2,(3,4)}$ | 0.00 | $\lambda_{15b,1,(3)}$ | 0.00 |
| $\lambda_{1,1,(4)}$ | 7.55* | $\lambda_{8a,2,(1,3)}$ | 0.36 | $\lambda_{10c,1,(2)}$ | 0.00 | $\lambda_{15b,2,(2,3)}$ | 0.00 |
| $\lambda_{1,2,(1,4)}$ | 0.00 | $\lambda_{8a,1,(2)}$ | 0.00 | $\lambda_{10c,2,(1,2)}$ | 1.05 | $\lambda_{15c,1,(1)}$ | 0.00 |
| $\lambda_{2,1,(1)}$ | 8.28* | $\lambda_{8a,2,(2,3)}$ | 0.01 | $\lambda_{10c,2,(2,4)}$ | 0.00 | $\lambda_{15c,2,(1,2)}$ | 0.00 |
| $\lambda_{2,2,(1,3)}$ | 6.93* | $\lambda_{8b,1,(1)}$ | 0.00 | $\lambda_{10c,1,(3)}$ | 0.02 | $\lambda_{15c,1,(3)}$ | 1.79 |
| $\lambda_{2,1,(2)}$ | 14.28** | $\lambda_{8b,2,(1,3)}$ | 0.00 | $\lambda_{10c,2,(1,3)}$ | 4.33* | $\lambda_{15c,2,(2,3)}$ | 0.00 |
| $\lambda_{2,2,(2,3)}$ | 0.29 | $\lambda_{8b,1,(2)}$ | 0.00 | $\lambda_{10c,2,(3,4)}$ | 0.10 | $\lambda_{16,1,(2)}$ | 0.00 |
| $\lambda_{2,1,(4)}$ | 4.82* | $\lambda_{8b,2,(2,3)}$ | 0.00 | $\lambda_{11,1,(2)}$ | 1.93 | $\lambda_{16,2,(1,2)}$ | 0.23 |
| $\lambda_{2,2,(3,4)}$ | 1.75 | $\lambda_{8b,1,(4)}$ | 0.00 | $\lambda_{11,2,(1,2)}$ | 2.38 | $\lambda_{16,1,(3)}$ | 0.00 |
| $\lambda_{3,1,(1)}$ | 23.35** | $\lambda_{8b,2,(3,4)}$ | 0.00 | $\lambda_{11,1,(3)}$ | 8.39* | $\lambda_{16,2,(1,3)}$ | 0.00 |
| $\lambda_{3,2,(1,2)}$ | 21.09** | $\lambda_{8c,1,(1)}$ | 2.42 | $\lambda_{11,2,(1,3)}$ | 1.37 | $\lambda_{16,1,(4)}$ | 0.73 |
| $\lambda_{3,1,(3)}$ | 6.08* | $\lambda_{8c,2,(1,3)}$ | 1.59 | $\lambda_{12,1,(2)}$ | 0.75 | $\lambda_{16,2,(1,4)}$ | 0.60 |
| $\lambda_{3,2,(2,3)}$ | 0.64 | $\lambda_{8c,1,(2)}$ | 3.52* | $\lambda_{12,2,(1,2)}$ | 0.03 | $\lambda_{17,1,(3)}$ | 0.92 |
| $\lambda_{3,1,(4)}$ | 2.10 | $\lambda_{8c,2,(2,3)}$ | 1.51 | $\lambda_{12,1,(3)}$ | 4.22* | $\lambda_{17,2,(2,3)}$ | 0.00 |
| $\lambda_{3,2,(2,4)}$ | 1.82 | $\lambda_{8c,1,(4)}$ | 1.39 | $\lambda_{12,2,(1,3)}$ | 3.79* | $\lambda_{17,1,(4)}$ | 2.81* |
| $\lambda_{4,1,(2)}$ | 0.10 | $\lambda_{8c,2,(3,4)}$ | 0.06 | $\lambda_{12,1,(4)}$ | 1.43 | $\lambda_{17,2,(2,4)}$ | 0.76 |
| $\lambda_{4,2,(1,2)}$ | 0.00 | $\lambda_{8d,1,(1)}$ | 19.37** | $\lambda_{12,2,(1,4)}$ | 1.39 | $\lambda_{18,1,(3)}$ | 4.21* |
| $\lambda_{4,1,(3)}$ | 2.47 | $\lambda_{8d,2,(1,3)}$ | 18.37** | $\lambda_{13,1,(1)}$ | 4.96* | $\lambda_{18,2,(1,3)}$ | 0.00 |
| $\lambda_{4,2,(1,3)}$ | 0.00 | $\lambda_{8d,1,(2)}$ | 20.99** | $\lambda_{13,2,(1,2)}$ | 4.89* | $\lambda_{18,2,(2,3)}$ | 0.00 |
| $\lambda_{4,1,(4)}$ | 1.03 | $\lambda_{8d,2,(2,3)}$ | 22.25** | $\lambda_{13,2,(1,4)}$ | 4.23* | $\lambda_{18,1,(4)}$ | 1.39 |
| $\lambda_{4,2,(1,4)}$ | 0.00 | $\lambda_{8d,1,(4)}$ | 5.78* | $\lambda_{13,1,(3)}$ | 3.95* | $\lambda_{18,2,(1,4)}$ | 0.00 |
| $\lambda_{5,1,(2)}$ | 3.97* | $\lambda_{8d,2,(3,4)}$ | 2.92* | $\lambda_{13,2,(2,3)}$ | 4.71* | $\lambda_{18,2,(2,4)}$ | 0.28 |
| $\lambda_{5,2,(1,2)}$ | 0.00 | $\lambda_{9,1,(2)}$ | 0.00 | $\lambda_{13,2,(3,4)}$ | 5.47* | $\lambda_{21,1,(2)}$ | 1.65 |
| $\lambda_{5,1,(3)}$ | 3.16* | $\lambda_{9,2,(1,2)}$ | 0.04 | $\lambda_{14,1,(3)}$ | 2.35 | $\lambda_{21,2,(1,2)}$ | 0.15 |
| $\lambda_{5,2,(1,3)}$ | 0.00 | $\lambda_{9,1,(3)}$ | 0.00 | $\lambda_{14,2,(1,3)}$ | 1.26 | $\lambda_{21,1,(3)}$ | 2.09 |
| $\lambda_{6,1,(1)}$ | 23.31** | $\lambda_{9,2,(1,3)}$ | 0.00 | $\lambda_{14,2,(2,3)}$ | 0.06 | $\lambda_{21,2,(1,3)}$ | 1.17 |
| $\lambda_{6,2,(1,2)}$ | 21.53** | $\lambda_{9,1,(4)}$ | 0.00 | $\lambda_{14,1,(4)}$ | 2.14 | $\lambda_{21,1,(4)}$ | 0.03 |
| $\lambda_{6,1,(3)}$ | 3.52* | $\lambda_{9,2,(1,4)}$ | 0.00 | $\lambda_{14,2,(1,4)}$ | 2.16 | $\lambda_{21,2,(1,4)}$ | 3.01* |
| $\lambda_{6,2,(2,3)}$ | 0.40 | $\lambda_{10a,1,(2)}$ | 0.90 | $\lambda_{14,2,(2,4)}$ | 0.96 | $\lambda_{22,1,(3)}$ | 1.99 |
| $\lambda_{6,1,(4)}$ | 12.72** | $\lambda_{10a,2,(2,4)}$ | 0.00 | $\lambda_{15a,1,(1)}$ | 5.72* | $\lambda_{22,2,(1,3)}$ | 0.71 |
| $\lambda_{6,2,(2,4)}$ | 9.37* | $\lambda_{10a,1,(3)}$ | 0.26 | $\lambda_{15a,2,(1,2)}$ | 4.76* | $\lambda_{22,2,(2,3)}$ | 0.00 |
| $\lambda_{7,1,(2)}$ | 0.00 | $\lambda_{10a,2,(3,4)}$ | 0.00 | $\lambda_{15a,2,(1,4)}$ | 8.17* | $\lambda_{22,1,(4)}$ | 3.96* |
| $\lambda_{7,2,(1,2)}$ | 0.03 | $\lambda_{10b,1,(2)}$ | 0.00 | $\lambda_{15a,1,(3)}$ | 1.24 | $\lambda_{22,2,(1,4)}$ | 0.41 |
| $\lambda_{7,1,(3)}$ | 0.00 | $\lambda_{10b,2,(1,2)}$ | 0.00 | $\lambda_{15a,2,(2,3)}$ | 2.24 | $\lambda_{22,2,(2,4)}$ | 0.15 |

*Note.* DTMR = Diagnosing Teacher's Multiplicative Reasoning; MI = modification index.
*$p < .05$  **$p < (.05/148)$

Table 8. DINA model modification indices for the DTMR fractions test data

| Parameter | MI | Parameter | MI | Parameter | MI |
| --- | --- | --- | --- | --- | --- |
| $\lambda_{5,1,(1)}$ | 3.21* | $\lambda_{13,1,(2)}$ | 0.00 | $\lambda_{17,1,(1)}$ | 0.00 |
| $\lambda_{5,1,(4)}$ | 0.00 | $\lambda_{13,1,(4)}$ | 1.19 | $\lambda_{17,1,(2)}$ | 8.08* |
| $\lambda_{8a,1,(3)}$ | 28.32** | $\lambda_{14,1,(1)}$ | 0.10 | $\lambda_{18,1,(1)}$ | 0.63 |
| $\lambda_{8a,1,(4)}$ | 0.00 | $\lambda_{14,1,(2)}$ | 0.00 | $\lambda_{18,1,(2)}$ | 5.63* |
| $\lambda_{10b,1,(1)}$ | 0.00 | $\lambda_{15a,1,(2)}$ | 0.00 | $\lambda_{22,1,(1)}$ | 1.67 |
| $\lambda_{10b,1,(4)}$ | 168.84** | $\lambda_{15a,1,(4)}$ | 0.45 | $\lambda_{22,1,(2)}$ | 6.89* |
| $\lambda_{10c,1,(1)}$ | 0.00 | $\lambda_{15b,1,(2)}$ | 2.90* | | |
| $\lambda_{10c,1,(4)}$ | 146.33** | $\lambda_{15b,1,(4)}$ | 0.00 | | |
| $\lambda_{11,1,(1)}$ | 0.05 | $\lambda_{15c,1,(2)}$ | 2.01 | | |
| $\lambda_{11,1,(4)}$ | 1.86 | $\lambda_{15c,1,(4)}$ | 0.05 | | |

*Note*. DINA = Deterministic Input Noisy And Gate; DTMR = Diagnosing Teachers' Multiplicative Reasoning; MI = modification index.
*$p < .05$  **$p < (.05/26)$